\documentstyle[twocolumn,aps,epsfig]{revtex}
\begin{document}
\wideabs{
\title{Interplay of linear and nonlinear impurities in the formation of \\
stationary localized states}

\author{Bikash C. Gupta and Sang Bub Lee}
\address{Department of Physics, Kyungpook National University,
Taegu 702--701, Korea}

\maketitle
\begin{abstract}
Formation of stationary localized states in one-dimensional chain 
as well as in a Cayley tree due to a linear impurity and a nonlinear 
impurity is studied.
Furthermore, a one-dimensional chain with linear and nonlinear 
site energies at the alternate sites is studied and
rich phase diagrams of SL states are obtained for all systems we considered.
The results are compared with those of the linear and nonlinear systems.   
\end{abstract}
}

\section{introduction}
The discrete nonlinear Schr\"odinger equation (DNLSE) describes a 
number of phenomena in condensed matter physics, 
nonlinear optics and other fields of physics\cite{scot,pete,emin,camp}.
The DNLSE in one dimension is written as
\begin{equation}
i \frac{d C_{n}}{dt} = \epsilon_{n} C_{n} + V (C_{n+1} + C_{n-1})
+ \chi_{n} f_{n} ( | C_{n} | ) C_{n} ,
\end{equation}
where $n=1,~2,~3, \cdots ,~ N$. 
The $\epsilon_{n}$ is the linear site energy at the site $n$, 
$V$ is the nearest-neighbor hopping matrix element, and $\chi_{n}$ 
is the nonlinear parameter associated with the $n$th site of the lattice.
Since $\sum_{n} | C_{n} |^{2}$ can be made unity by 
choosing appropriate initial conditions, 
$| C_{n} |^{2}$ may be considered to be the probability of 
finding a particle at the $n$th site in the lattice. 
Consequently, $C_{n}$ may be interpreted as the probability amplitude 
of a particle at the $n$th site.
In the context of the condensed matter physics, 
the nonlinear term $f_{n} (| C_{n} |)$ arises from the 
coupling of vibrations of masses at the lattice point to the motion of a 
quasiparticle in the limit of high frequency lattice vibrations. 
If the lattice oscillators are purely harmonic in nature 
and are independent of one another, it follows that 
$f_{n} (| C_{n} |) = |C_{n}|^{2}$\cite{chen,wany,mol1}. 
As an application of DNLSE particularly in condensed matter physics, 
we cite among others the exciton propagation in 
Holstein molecular chain\cite{scot}. 
In general, the exciton propagation in quasi one dimension
having short range electron-phonon interaction can be adequately modeled
by the DNLSE\cite{cohe}. 
Other examples include nonlinear optical responses in superlattices 
formed by dielectric or magnetic slabs\cite{mill} 
and the mean field theory of aperiodic array of twinning planes 
in the high $T_{c}$ superconductors\cite{abri}. 

One important feature of the DNLSE is that it can produce stationary
localized (SL) states. 
The SL states might play an important role in the nonlinear 
DNA dynamics\cite{daux} and also in energy localization 
in nonlinear lattices\cite{caid}. 
It has been shown that the presence of impurities can produce SL states 
in one, two and three 
dimensions\cite{mol2,mol3,hui1,hui2,hui3,acev,wein,bik1,bik2,kun,bik3,bik4}.
The formation of SL states in one-dimensional chain with a single nonlinear
impurity and that with a dimeric nonlinear impurity have been studied in
detail\cite{mol2,bik1}. 
The same problem has also been studied starting from an 
appropriate Hamiltonian\cite{wein,bik2}.
The fixed point of the Hamiltonian which generates the appropriate DNLSE 
can also produce the correct equation, giving the formation of SL states.
The formation of SL states in a perfect nonlinear chain containing 
a single nonlinear impurity and also in a perfect nonlinear chain with
a dimeric nonlinear impurity has been studied\cite{kun}.  
The formation of SL states in a perfectly nonlinear Cayley tree 
due to various nonlinear impurities has also been 
considered earlier\cite{bik1,bik3,bik4}. 

On the other hand, there has been a detailed study on the formation of 
localized states due to linear impurities in one-dimensional chain as 
well as in the Cayley tree\cite{econ}. 
The number of localized states do not exceed the number of impurities 
in case of linear systems.
We note that the stationary localized states have appeared due to the 
presence of either linear impurities or nonlinear impurities.
Thus, one may naturally ask the question of what happens when
both the linear and nonlinear impurities are present 
at the same time in the system.
In order to investigate the mixed role of linear and nonlinear 
impurities in the formation of SL states, 
we consider a simple one-dimensional chain as well as a Cayley tree 
consisting of a single linear impurity and a single nonlinear 
impurity. 
Furthermore, we consider a one-dimensional chain with linear and 
nonlinear site energies at alternate sites of the lattice.

This paper is organized as follows. 
The one-dimensional chain with a linear impurity and a 
nonlinear impurity is considered in Sec.~II.
In Sec.~III, we consider the one-dimensional chain with linear and 
nonlinear site energies at alternate sites. 
The Cayley tree with a linear impurity and a nonlinear impurity is 
considered in Sec.~IV.  
We finally summarize our observations in Sec.~V.

\section{One-dimensional chain with linear and nonlinear impurities}
Consider a one-dimensional chain consisting of a linear impurity of strength
$\epsilon$ placed at the zeroth site and a nonlinear impurity of
strength $\chi$ at the first site.
The relevant Hamiltonian for such a system can be written as
\begin{equation}
H = \frac{1}{2} \sum_{n= - \infty}^{\infty} 
     (C_{n} C_{n+1}^{\star} + C_{n}^{\star} C_{n+1}) 
   + \frac{\epsilon}{2} | C_{0} |^{2} + \frac{\chi}{4} |C_{1}|^{4} ,
\end{equation}
where the nearest-neighbor hopping matrix element is assumed to be unity.
As a possible solution for the stationary localized states,
we consider
\begin{equation}
C_{n} = \phi_{n} e^{-iEt} ,
\end{equation}
where $E$ is the stationary state energy. 
It is also well known that impurity states in one-dimensional system
are exponentially localized. 
Therefore, the presence of two impurities at the two consecutive sites 
(site 0 and site 1) in the chain, 
suggests us to consider the following forms of $\phi_{n}$: 
\begin{eqnarray}
\phi_{n}   & = & \phi_{1} \left[ {\rm sgn}(E) \eta \right]^{n-1} ~~~~~ 
n \ge 1 \nonumber \\
\phi_{-|n|} & = & \phi_{0} \left[ {\rm sgn}(E) \eta \right]^{|n|} 
~~~~~~~ n \le 0 . 
\end{eqnarray}
The form of $\phi_{n}$ given above are exact and may be derived from the
Green's function analysis\cite{bik1}. 
Here $\eta$ lies between 0 and 1 and is given as
\begin{equation}
\eta = \frac{1}{2} \left[ |E| - \sqrt{E^{2} - 4} \right] .
\end{equation}
Since we are dealing with the localized states, $\phi_{0}$ and $\phi_{1}$ 
can be assumed to be real without forsaking the mathematical rigor. 
Three possible cases may be considered, i.e., 
$\phi_{1} = \phi_{0}$, $\phi_{1} = - \phi_{0}$, and $\phi_{1} \ne \phi_{0}$,
the solutions for which are called, respectively, symmetric, 
antisymmetric, and asymmetric solutions. 
To take into account of all three possibilities, 
we introduce a variable $\beta = \phi_{0} / \phi_{1}$. 
Consequently, $\beta = {1}$ corresponds to the symmetric solutions, 
$\beta = -1$ to the antisymmetric solutions, and $\beta \ne \pm 1$
to the asymmetric solutions. 
Substituting the ansatz of Eq.~(4) and the definition of $\beta$ 
to the normalization condition $\sum | C_{n} |^{2} = 1$, we obtain 
\begin{equation}
| \phi_{0} |^{2} = \frac{1- \eta^{2}}{1 + \beta^{2}}.
\end{equation}
Using Eqs.~(3), (4) and (6) in Eq.~(2),
we obtain the effective Hamiltonian for the reduced dynamical system:
\begin{equation}
H_{\rm eff} = {\rm sgn}(E) \eta + \frac{(1 - \eta^{2} )}{(1 + \beta^{2})}
\left( \beta + \frac{\epsilon}{2} + \frac{\chi}{4} 
\frac{\beta^{4} (1 - \eta^{2})}{(1 + \beta^{2})} \right) .
\end{equation}

We first consider the case of $\beta = \pm 1$. 
The effective Hamiltonian for this case becomes
\begin{equation}
H_{\rm eff}^{\pm} = {\rm sgn}(E) \eta \pm \frac{(1 - \eta^{2})}{2} 
+ \frac{\epsilon (1 - \eta^{2})}{4} 
+ \frac{\chi}{16} \beta^{4} (1 - \eta^{2} )^{2} ,
\end{equation}
where ``$+$'' refers to the symmetric states and ``$-$'' corresponds to the
anti-symmetric states. 
For a particular system, $\epsilon$ and $\chi$ are constants,
while $\eta$ is a variable which determines the energy of the localized state. 
The value of $\eta$ should be determined self-consistently and, 
for this reason, we treat $\eta$ as the dynamic variable in the 
effective Hamiltonian of the reduced dynamical system. 
The fixed-point solutions corresponding to the localized states 
can be found, from the variabtional principle, by the condition
$\partial H_{\rm eff}^{\pm}/ \partial \eta = 0$ with $\eta \in [0,1]$.
Since the form of the probability profile is rigorous\cite{econ}, 
we obtain the correct values of $\eta$, satisfying the equation 
\begin{equation}
\frac{4}{\chi^{\pm}} = 
\frac{\eta (1 - \eta^{2})}{sgn(E) - ( \frac{\epsilon}{2} \pm 1) \eta} 
= f_{\pm}(\eta).
\end{equation}
>From Eq.~(9) it is clear that the states will appear above the band for
$\chi^{\pm} > 0$ and below the band for $\chi^{\pm} < 0$.  
[Note that the band edges are at $E/V = \pm 2$.]
We will consider positive values of $\chi^{\pm}$ and, hence, the 
sgn($E$) may be taken to be positive. 
For symmetric states, 
the value of $f_{+} ( \eta )$ exhibits the extreme values:
$f_{+} ( \eta ) \rightarrow 0$ as $\eta \rightarrow 0$ and 
$f_{+} ( \eta ) \rightarrow \infty$ as 
$\eta \rightarrow 1 / ( \epsilon /2 + 1 )$. 
Thus, $f_{+} ( \eta )$ increases as $\eta$ increases from $\eta = 0$ 
and diverges at $\eta = 1 / ( \epsilon /2 + 1)$, 
and it becomes negative for $\eta > 1 / ( \epsilon /2 + 1)$. 
This means that there will always be only one solution (symmetric state) 
for all positive values of $\epsilon$ and $\chi$. 
For antisymmetric states, we found that 
$f_{-} ( \eta ) \rightarrow 0$ as $\eta \rightarrow 0$ and
$f_{-} ( \eta ) \rightarrow \infty$ as 
$\eta \rightarrow 1 / ( \epsilon / 2 - 1)$.
The divergence of $f_{-} ( \eta )$ is, however, confined between 0 and 1 
if and only if $\epsilon > 4$. 
Therefore, for $\epsilon > 4$, $f_{-} ( \eta )$ increases
as $\eta$ increases from $\eta = 0$, 
diverges at $\eta = 1 / ( \epsilon /2 - 1)$, and finally becomes negative 
for $\eta > 1/ ( \epsilon /2 - 1)$. 
This implies that there will always be one antisymmetric state 
for $\epsilon > 4$ for any value of $\chi > 0$. 
However, the situation is different for $\epsilon < 4$;
the value of $f_{-} ( \eta )$ approaches 0 as $\eta \rightarrow 0$ 
and $\eta \rightarrow 1$ and remains positive and finite for $0< \eta <1$.
This implies that $f_{-} ( \eta )$ has at least one maximum in the range of 
$0 < \eta <1$. 
We confirm by a graphical analysis that there is only one maximum. 
Therefore, there will be a critical value of $\chi$ separating 
two antisymmetric states from the region of no antisymmetric state 
for $\epsilon < 4$. 
The solid line in Fig.~1 is such a critical line and is denoted by 
$\chi_{cr}^{a}$.
Thus, there are three regions in the entire plane of $\epsilon$ versus 
$\chi$, having different number of symmetric and antisymmetric SL states. 
The region of $\epsilon > 4$ has one symmetric and one antisymmetric SL
states, while the region of $\epsilon < 4$ is divided into two regions 
by the critical line $\chi_{cr}^{a}$, 
below of which has only one symmetric state and above of which has 
one symmetric state and two antisymmetric states. 
Therefore, the maximum number of SL states contributed by the symmetric and 
anti-symmetric solutions is three.

Next, we consider the asymmetric solutions, i.e., for $\beta \ne \pm 1$. 
The effective Hamiltonian in Eq.~(6) should be considered as a function of 
two variables, $\eta$ and $\beta$, 
while $\epsilon$ and $\chi$ remain constant for a particular system. 
The fixed-point solutions corresponding to the SL states
can be obtained from the following two extreme conditions
\begin{eqnarray}
\partial H_{\rm eff} / \partial \eta & = & 0 \nonumber \\
\partial H_{\rm eff} / \partial \beta & = & 0 .
\end{eqnarray}
After a trite algebra, the fixed-point equations reduce to
\begin{eqnarray}
( 1 + \beta^{2} ) ( 1 + \beta^{2} - 2 \eta \beta  - \epsilon \eta )
- \chi \eta \beta^{4} (1- \eta^{2}) & = & 0 \nonumber \\
(1 - \beta^{4} ) - \epsilon \beta (1 + \beta^{2}) 
  + \chi \beta^{3} (1 - \eta^{2} ) & = & 0.
\end{eqnarray}
These coupled equations can not be solved analytically and, accordingly,
we analyze them numerically to find the number of asymmetric SL states 
(number of solutions with $\eta \in [0,1]$). 

\begin{center}
\epsfig{file=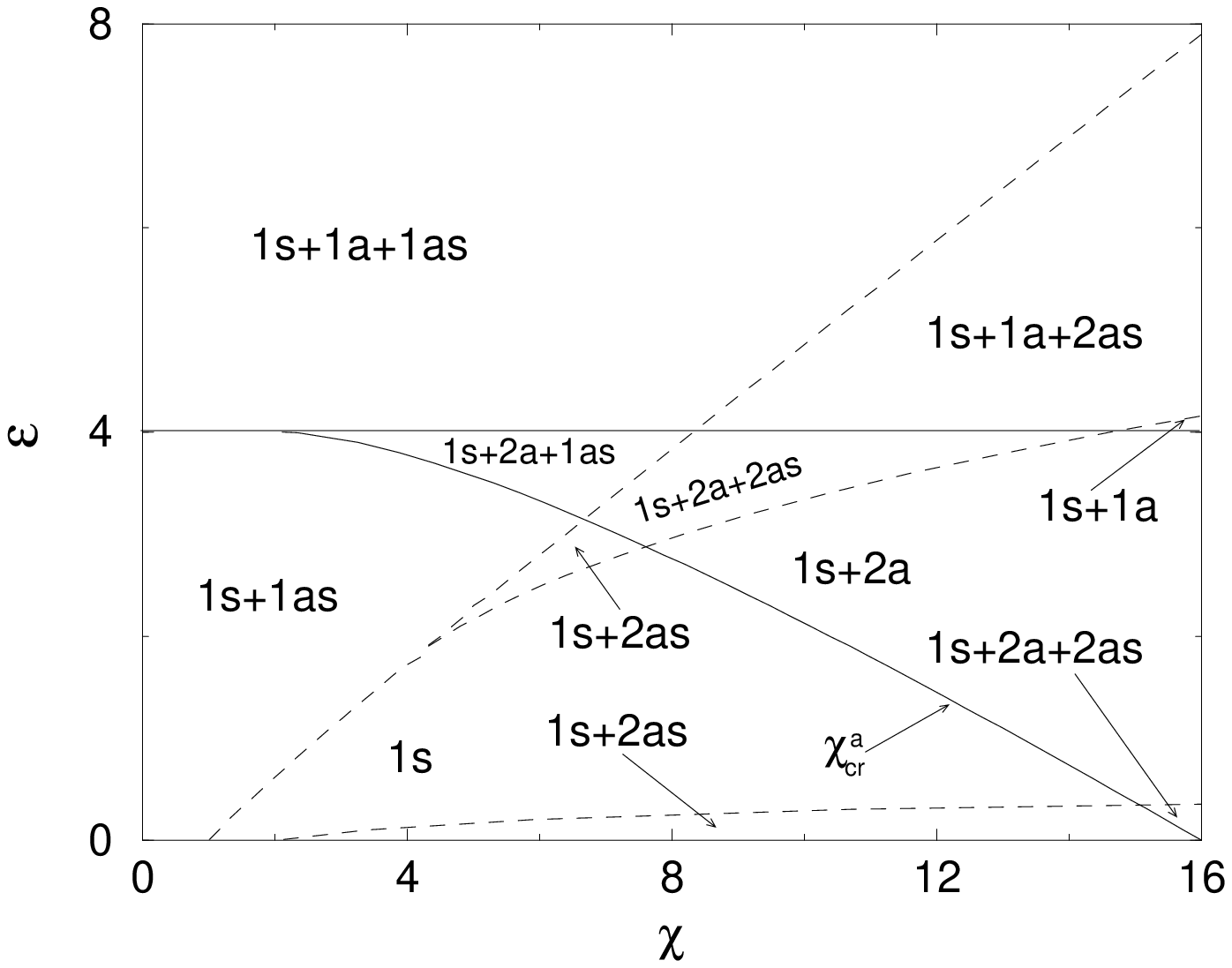,width=7.8cm,height=6.2cm}
\begin{figure}[bh]
\caption{The phase diagram for SL states in the plane of 
         $\epsilon$ versus $\chi$ for a system of a one-dimensional chain 
         with a linear impurity and a nonlinear impurity. 
         Different regions are marked with the number of states 
         and the kind of states; the region marked by ``1s+2a+2as'' means 
         that the region has one symmetric, 
	 two antisymmetric and two asymmetric SL states.} 
\label{fig1}
\end{figure}
\end{center}
\vspace{0.2cm}

It is found that there are three critical lines separating 
different number of states in the plane of $\epsilon$ vs $\chi$
denoted by the dashed curves in Fig.~1. 
Maximum number of asymmetric SL states obtained is two.
Considering all possible states we find that maximum number 
of SL states is five. 
The ``s'', ``a'' and ``as'' stand for symmetric, antisymmetric and 
asymmetric states, respectively. 
For example, the region marked as ``1s+2a+2as'' has one symmetric SL state, 
two antisymmetric SL states, and two asymmetric SL states. 
It should be noted that the number of states changes one by one as 
we go from one region to another in the phase space. 
As an example, as we cross from the region ``1s+1as'' to the region
``1s+2a+1as'', the number of states increases by two;
however, the number of states on the critical line which separates
two regions is ``1s+1a+1as'', yielding the increment of the number of
states one by one.
The maximum number of SL states with two nonlinear impurities 
was also found to be five\cite{bik3}. 
We thus conclude that one linear impurity and one nonlinear impurity play
a similar role to the two nonlinear impurities as far as the 
maximum number of SL states are concerned.
However, the phase diagrams for two cases are different. 

Figure~2 shows the variation of energy of the states 
as a function of $\chi$ for a fixed value of $\epsilon = 2$. 
As $\chi$ increases, the energy of the state increases 
(goes away from the band edge) and, beyond $\chi \approx 10.39$, 
two more states appear as expected from the phase diagram in Fig.~1. 
The energy of one of the two new states increases and that of the rest 
decreases and approaches to the band edge as $\chi$ increases.
Therefore, for $\chi < 10.39$, the localization length of the state, i.e., 
the characteristic length of the profile, decreases as $\chi$ increases, 
whereas, for $\chi > 10.39$, the localization length of one of 
the three states increases as $\chi$ increases. 
Thus, for very large $\chi$, two of the states are strongly localized and 
the rest is weakly localized. 

\vspace{0.4cm}
  \begin{center}
  \epsfig{file=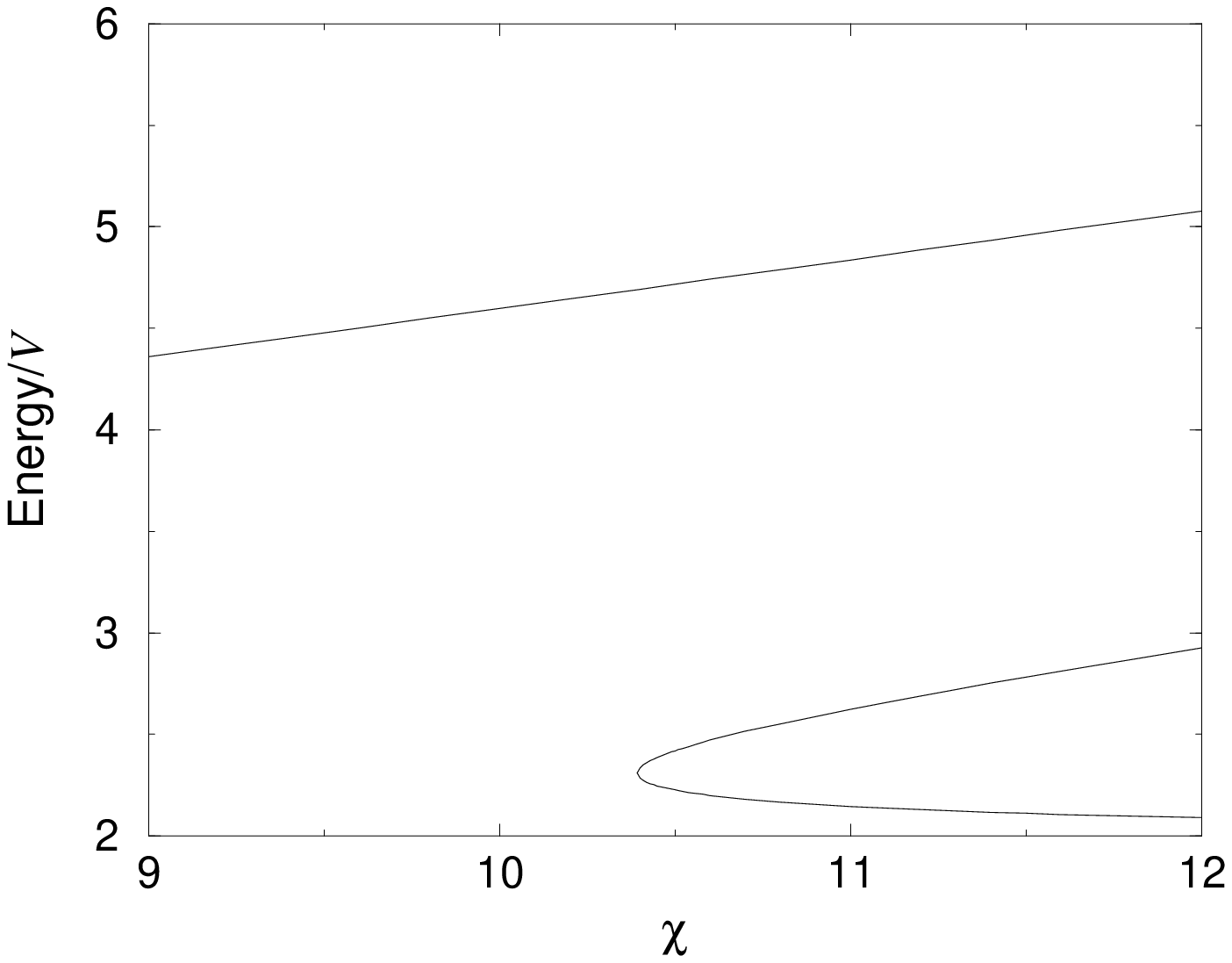,width=7.4cm,height=5.7cm}
  \begin{figure}[bh]
  \caption{Variation of energy as a function of $\chi$ is shown for 
           $\epsilon = 2$.}
  \label{fig2}
  \end{figure}
  \end{center}

\section{Linear and nonlinear site energies at alternate sites}
In this section, we consider a one-dimensional chain with alternate 
site energies; the even sites have a linear energy $\epsilon$ and 
the odd sites have a nonlinear energy $\chi |C_{n}|^{2}$. 
Such a system can be described by the Hamiltonian given as
\begin{eqnarray}
H & = & \frac{1}{2} \sum_{n= - \infty}^{\infty} 
        ( C_{n} C_{n+1}^{\star} + C_{n}^{\star} C_{n+1} ) \nonumber \\
  &   & + \frac{\epsilon}{2} \sum_{n= - \infty}^{\infty} | C_{2n}|^{2}
        + \frac{\chi}{4} \sum_{n= - \infty}^{\infty} |C_{2n+1}|^{4} .
\end{eqnarray}
For $\chi = 0$, the system reduces to a periodically modulated linear system. 
The translational symmetry of the systems are preserved for both
$\chi = 0$ and $\chi \ne 0$. 
Using $C_{n} = \phi_{n} \exp (- iEt)$, 
we obtain the Hamiltonian in terms of $\phi_{n}$. 
It is, however, not possible to find the exact ansatz for the localized
states in this system, although a certain rational ansatz can be considered.
For example, the onsite peaked solution, intersite peaked or intersite
dipped solutions are possible. 
We will consider each of these in the subsequent subsections.

\subsection{Onsite peaked solution}
Since the system has linear and nonlinear sites at regular intervals,
the solution may have a peak either at the linear site or at the nonlinear
site.
We first consider the solution peaked at the linear site, i.e. the
solution is peaked at the central site of the system. 
We therefore consider the ansatz $\phi_{n} = \phi_{0} \eta^{|n|}$. 
Using this and the normalization condition,
we obtain the effective Hamiltonian as
\begin{equation}
H_{\rm eff} = \frac{2 \eta}{1 + \eta^{2}} + \frac{\epsilon}{2} 
\frac{ (1 + \eta^{4}) }{ (1 + \eta^{2} )^{2}} + \frac{\chi}{2}
\frac{\eta^{4} (1 - \eta^{2} )^{2} }{ (1 + \eta^{2} )^{2} (1 - \eta^{8} )}.
\end{equation}
The fixed-point equation, $\partial H_{\rm eff} / \partial\eta = 0$, 
reduces to  
\begin{eqnarray}
&2&(1 - \eta^{8})^{2} (1 - \eta^{4}  - \epsilon \eta + \epsilon \eta^{3} ) 
  - 2 \chi \eta^{5} (1 - \eta^{2} )^{2} (1 - \eta^{8} ) \nonumber \\
&+& 2 \chi \eta^{3} (1 - \eta^{4} ) \left( ( 1 - 2 \eta^{2} ) (1 - \eta^{8} ) 
  + 2 \eta^{8} (1 - \eta^{2}) \right) = 0 \nonumber \\
\end{eqnarray}
Solving Eq.~(14), we found that there are two separate regions in the 
plane of $\epsilon$ versus $\chi$: one with three SL states another  
with only one SL state, as shown in Fig.~3.

\vspace{0.5cm}
  \begin{center}
   \epsfig{file=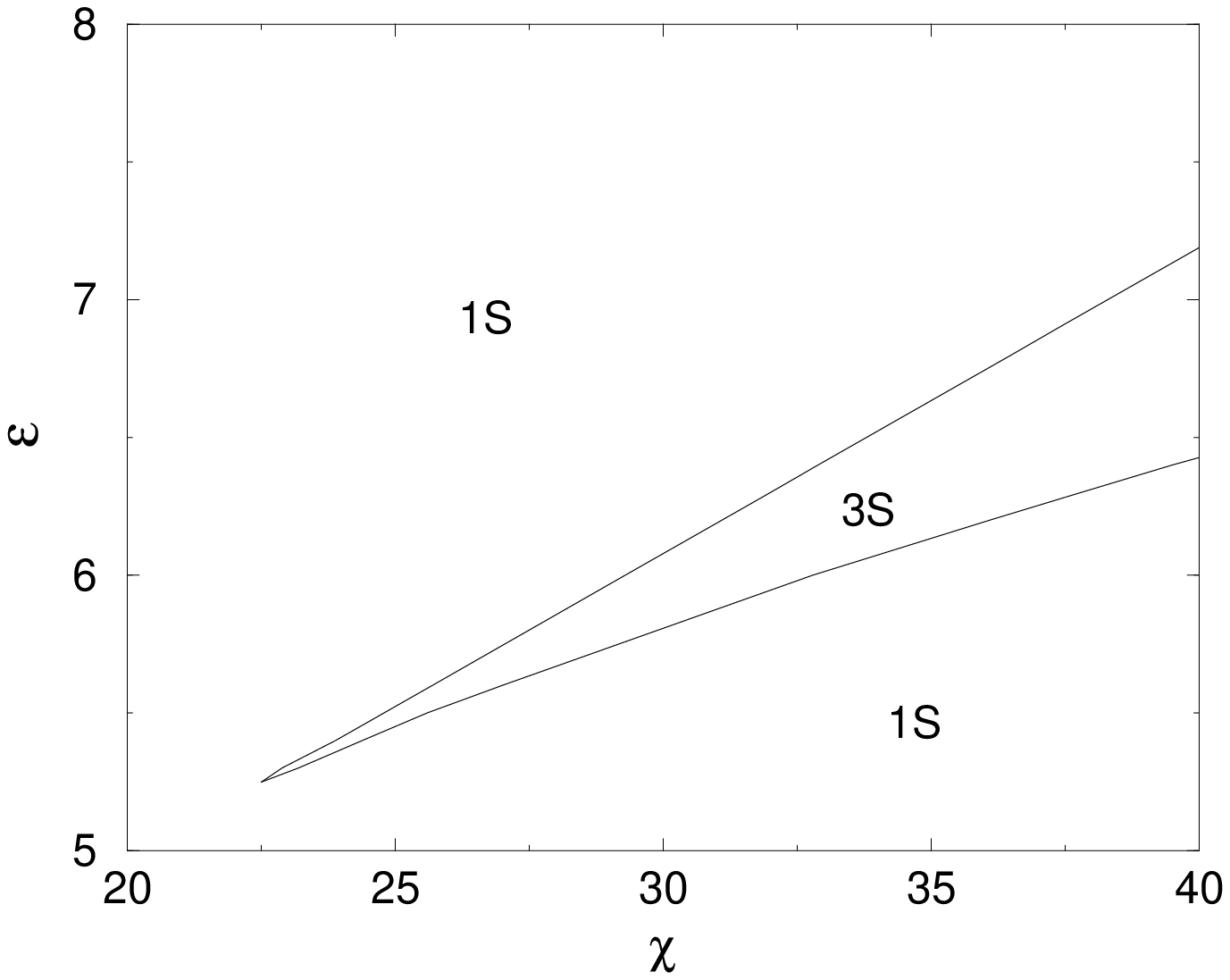,width=7.8cm,height=5.8cm}
   \begin{figure}[bh]
   \caption{The phase diagram for the SL states (solutions peaked at the 
            linear site) in the plane of $\epsilon$ versus $\chi$ for a  
	    one-dimensional chain with linear and nonlinear site
	    energies at alternate sites.}
   \label{fig3}
  \end{figure}
  \end{center}
\vspace{0.3cm}

The localized solution peaked at the nonlinear site may be obtained
considering that each site is shifted by one lattice site along
either positive or negative direction, and the resulting
fixed-point equation is give as
\begin{eqnarray}
& 2 & (1 - \eta^{2} ) ( 1- \eta^{8})^{2} ( 1+ \eta^{2} + \epsilon \eta) \nonumber \\
& - & \chi \eta (1- \eta^{4} ) (1 - \eta^{16} ) 
      + \chi \eta (1 - \eta^{4} ) (4 \eta^{6} ( 1 - \eta^{2} ) \nonumber \\
& - & ( 1 - \eta^{2} )^{2} (1 + \eta^{8} )( 1 + \eta^{4} ) ) = 0 .
\end{eqnarray}
The phase diagram for the SL states is presented in Fig.~4.
We again find that there are two separate regions,
similar to the case peaked at the linear site.
Thus, at least two onsite peaked SL states 
(one of them is peaked at the linear site and another peaked at the 
nonlinear site) may be obtained for any finite values of $\chi$ and $\epsilon$.
However, the regions of three SL states peaked at the nonlinear site 
appears from much lower values of $\chi$ 
and $\epsilon$, compared to the case peaked at the linear site.

\vspace{0.5cm}
  \begin{center}
  \epsfig{file=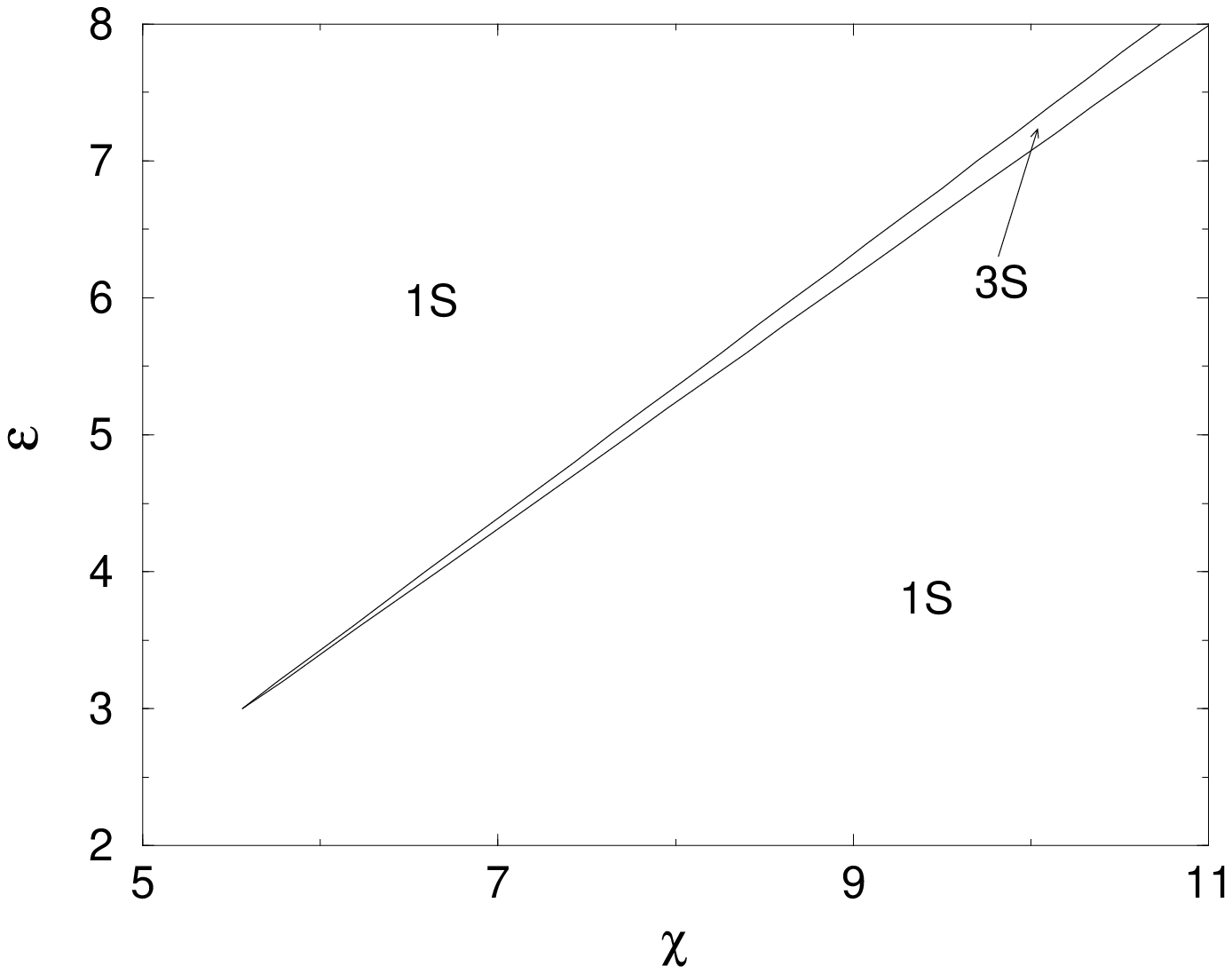,width=7.8cm,height=5.9cm}
  \begin{figure}[bh]
   \caption{The phase diagram for the SL states  
            (solutions peaked at the nonlinear site) in the plane of
	    $\epsilon$ versus $\chi$ for the system of a 
	    one-dimensional chain with linear and nonlinear site energies 
	    at alternate sites.}
   \label{fig4}
  \end{figure}
  \end{center}

\subsection{intersite peaked and dipped solutions}
For the intersite peaked and dipped solutions we use the dimeric ansatz
in Eq.~(4). 
After short calculatons, we obtain the effective Hamiltonian for the 
reduced dynamical system as
\begin{eqnarray}
H_{eff} & = & sgn(E) \eta 
    + \frac{\beta (1- \eta^{2} )}{1+ \beta^{2} } \nonumber \\
    & + & \frac{\epsilon}{2} \frac{ (1+ \beta^{2} \eta^{2} ) }{ (1 + \beta^{2} )
      ( 1+ \eta^{2} )} 
      +\frac{\chi}{4} \frac{ (1- \eta^{2} )^{2} ( \eta^{4} + \beta^{4} )}
    { ( 1+ \beta^{2} )^{2} (1- \eta^{8} )}
\end{eqnarray}
with $\beta$ and $\eta$ as defined earlier. 
We first consider the symmetric and antisymmetric states, i.e., $\beta = \pm1$. 
The fixed-point equation yields 
\begin{equation}
\frac{4}{\chi^{\pm}} = \frac{\eta}{(1 \mp \eta)( 1 + \eta^{2} )^{2} } 
= f^{\pm} ( \eta ) ,
\end{equation}
where ``+'' referes to the symmetric states and ``-'' to the 
antisymmetric states.  
It should be noted that $\epsilon$ has no role in the formation of 
symmetric and antisymmetric SL states in the system. 
>From Eq.~(17) it is clear that only one symmetric state appears for any 
finite value of $\chi$. 
However, for antisymmetric states, the situation is different.
The value of $f^{-} ( \eta )$ in Eq.~(17) initially increases 
as $\eta$ increases and finally decreases as $\eta$ approaches to 1,
leaving a maximum at $\eta \in [0,1]$. 
Thus, there are two critical values of $\chi$ such that there
is no anti-symmetric SL state for $\chi < 18.72$, 
two anti-symmetric SL states for $18.72 < \chi < 32$ 
and one anti-symmetric SL state for $\chi \ge 32$, 
shown as the two vertical lines in Fig.~5. 

We next consider the asymmetric states of $\beta \ne 1$. 
In this case $\epsilon$ plays its role, unlike the case of symmetric and
antisymmetric states.
The fixed-point equations are given as 
\begin{eqnarray}
&   & (1 + \eta^{2} ) ( 1 + \beta^{2} )( 1 - \eta^{8} )^{2} 
    (1 + \beta^{2} - 2 \beta \eta )  \nonumber \\
& - & \epsilon \eta (1 - \beta^{4} ) (1 - \eta^{8} )^{2} 
    - \chi \eta (1- \eta^{4} )^{2}  \nonumber \\
& \times & \left( ( \eta^{4} + \beta^{4} ) 
    (1 + \eta^{2} + \eta^{4} - \eta^{6} ) - \eta^{2} (1- \eta^{8} ) \right) = 0 
\end{eqnarray}
and
\begin{eqnarray}
& &(1- \eta^{4} )(1 + \beta^{2} )( 1- \eta^{8} )(1- \beta^{2} ) \nonumber \\
&-& \epsilon \beta (1 + \beta^{2} ) ( 1- \eta^{8} ) (1 - \eta^{2} ) \nonumber \\
&+& \chi \beta (1 - \eta^{2} ) (1- \eta^{4}) ( \beta^{2} - \eta^{4} ) = 0 .
\end{eqnarray}
Solving Eqs.~(18) and (19) we obtain various critical lines separating
the $\chi - \epsilon$ plane into various regions with various number
of asymmetric SL states as shown in Fig.~5. 
Complete phase diagram will be obtained by superposing the Figs.~3, 4, and 5. 
The figures are shown separately because otherwise it will appear very
complicated.
The maximum number of states obtained in this case is 9. 
We note that the presence of both linear and nonlinear sites
increases the number of SL states compared to the fully nonlinear case.
No SL state is obtained for $\chi = 0$. 
This is expected because the system with $\chi = 0$ reduces to a
periodically modulated linear system which again may be renormalized 
to a perfect linear system.

\vspace{0.5cm}
  \begin{center}
   \epsfig{file=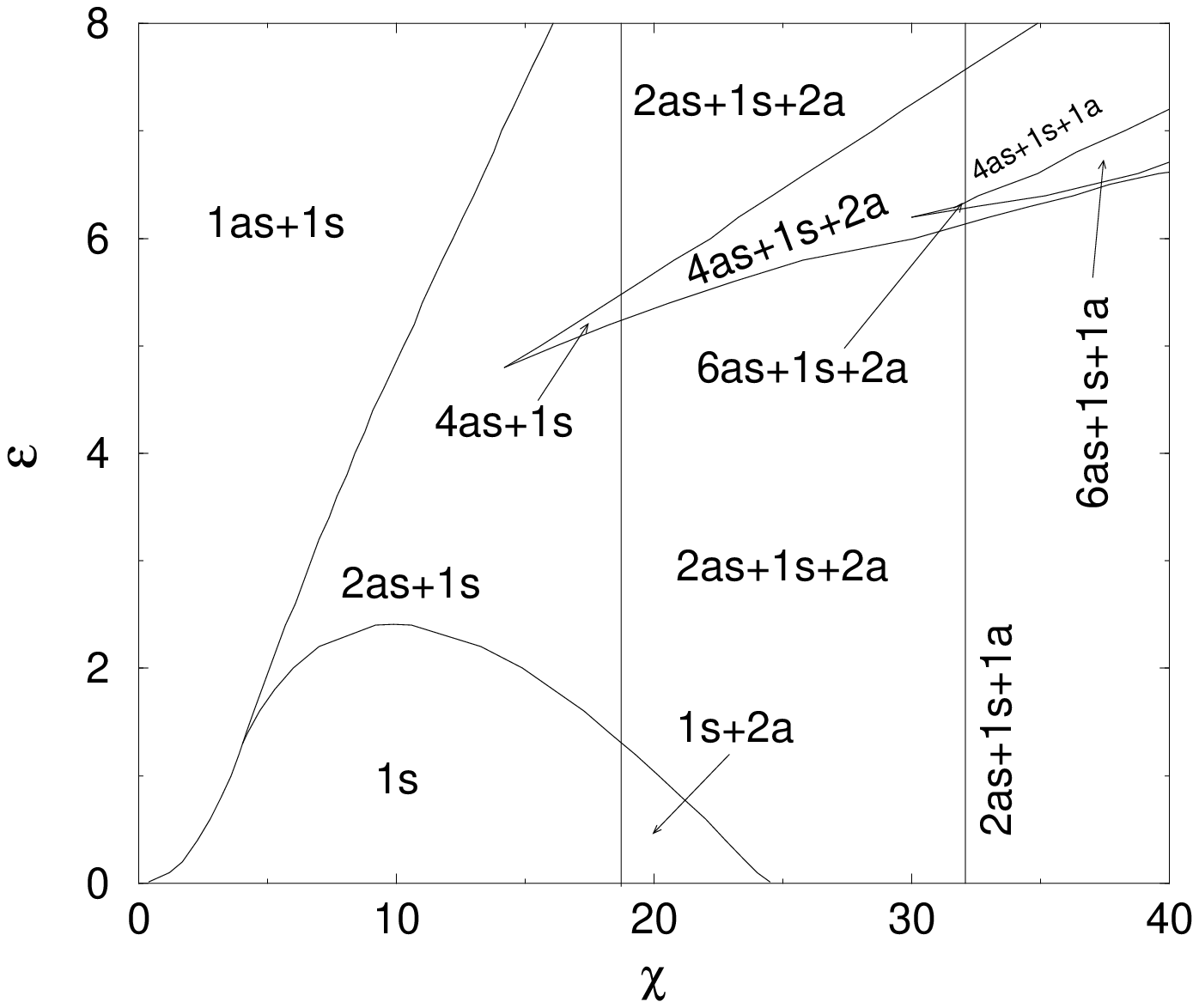,width=7.8cm,height=6.8cm}
   \vspace{-1.3cm}
   \begin{figure}[bh]
   \caption{The phase diagram for the SL states due to the intersite 
   peaked and dipped solutions in the plane of $\epsilon$ versus $\chi$ 
   for the  one-dimensional chain with a linear and a nonlinear 
   site energies at alternate sites.}
  \label{fig5}
  \end{figure}
  \end{center}

\section{Cayley tree with a linear impurity and a nonlinear impurity}
Here we consider a Cayley tree with a linear impurity and 
a nonlinear impurity embedded in it. 
The purpose of this work is to observe how the SL states vary according to
the geometry of the underlying lattice structure.
The structure of the Cayley tree for the coordination number $Z = 3$
and the connectivity $K = 2$ is shown in Fig.~6. 

\vspace{0.5cm}
  \begin{center}
  \epsfig{file=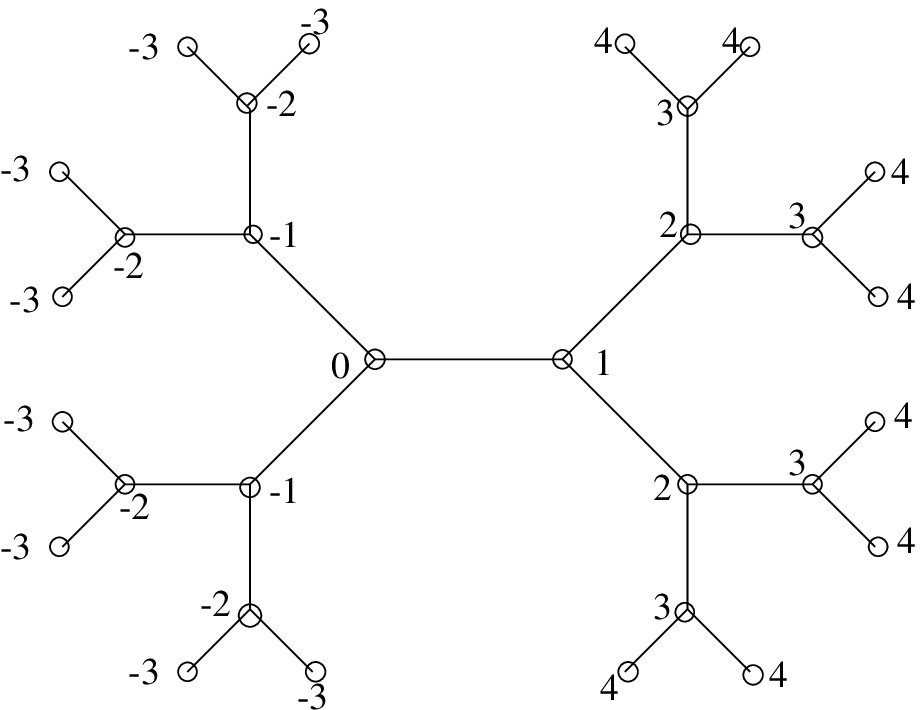,width=7.0cm,height=5.8cm}
  \vspace{0.2cm}
   \begin{figure}[bh]
   \caption{The structure of Cayley tree with the coordination 
            number $Z=3$ and the connectivity $K=2$.}
   \label{fig6}
  \end{figure}
  \end{center}

We pick a bond and assign the numbers 0 and 1 on its two ends.
The neighboring sites of the site 1 are numbered with an increasing order
and those of the site 0 are numbered with a decreasing order.
Thus, all points with the same number are assumed to be in the same generation.
The number of points in the $n$th generation is
$K^{n-1}$ if $n \ge 1$ and $K^{ |n| }$ if $n \le 0$.
We note that all sites in a given generation have the same
probability amplitude.

We now consider the motion of a particle on a Cayley tree of the
connectivity $K$ with two impurities ($\epsilon_{0}$ and $\epsilon_{1}$)
embedded at sites 0 and 1, respectively.
In the tight-binding formalism with nearest-neighbor hopping only,
the equations governing the motion of a particle are
\begin{eqnarray}
i \frac{d C_{n}}{dt} & = & K  C_{n+1} + C_{n-1}, ~~~~~~~~~~~ n > 1 
\nonumber \\
i \frac{d C_{n}}{dt} & = & K C_{-|n|-1} + C_{-|n|+1},~~~~~ n < 0 
\nonumber \\
i \frac{d C_{1}}{dt} & = & K C_{2} + C_{0} + \epsilon_{1} C_{1} 
\nonumber \\ 
i \frac{d C_{0}}{dt} & = & K C_{-1} + C_{1} + \epsilon_{0} C_{0} ,
\end{eqnarray}
where $C_{n}$ denotes the probability amplitude at any point in the $n$th 
generation.
We assume that the matrix element of the nearest-neighbor hopping is unity
and that all points in a given generation have the same site energy.
The normalization condition for the site amplitudes becomes
\begin{equation}
\sum_{n= - \infty}^{0} K^{|n|} |C_{n}|^{2} + \frac{1}{K} 
\sum_{n=1}^\infty K^{n} |C_{n}|^{2} = 1.
\end{equation}

Solving the Eq.~(20) appears to be nontrivial; however, by a simple
transformation of variables (i) $\tau = \sqrt{K} t$,
(ii) $\epsilon_{0} = \widetilde \epsilon_{0} / \sqrt{K}$ 
and $\epsilon_{1} = \widetilde \epsilon_{1} / \sqrt{K}$, and
(iii) $C_{n} = K^{-(n-1)/2} \widetilde C_{n}$ for $n \ge 1$ 
and $C_{-|n|} = K^{-|n|/2} \widetilde C_{n}$ for $n \le 0$
in Eq.~(20), we obtain
\begin{eqnarray}
i \frac{d \widetilde C_{n}}{d \tau} 
& = & \widetilde C_{n+1} + \widetilde C_{n-1} , ~~\mbox{for}~~ n >1 
~~\mbox{and}~~ n < 0. \nonumber \\
i \frac{d \widetilde C_{1}}{d \tau}
& = & \widetilde C_{2} + \frac{1}{\sqrt{K}}
\widetilde C_{0} + \widetilde \epsilon_{1} \widetilde C_{1} , \nonumber \\ 
i \frac{d \widetilde C_{0}}{d \tau}
& = & \widetilde C_{-1} + \frac{1}{\sqrt{K}}
\widetilde C_{1} + \widetilde \epsilon_{0} \widetilde C_{0} . 
\end{eqnarray}
which are precisely the equations for a one-dimensional chain,
with the normalized probability amplitude $\widetilde C_{n}$.
The normalization condition in Eq.~(21) reduces to
$\sum_{-\infty}^{\infty} | \widetilde C_{n} |^{2} = 1 $.
Therefore, the motion of a particle on a Cayley tree is mapped onto
that on a one-dimensional chain with a bond defect between the impurity sites. 
In other words, the nearest-neighbor hopping matrix element 
between the zeroth and the first site is reduced from unity 
to $1 / \sqrt{K}$.
It can be shown that the Green's function $G_{0,0} (E)$ calculated from Eq.~(7)
would yield $\widetilde G_{0,0} ( \widetilde E = E \sqrt{K})$
for a Cayley tree of the connectivity $K$.
Here, we consider a linear impurity at the zeroth site and a nonlinear
impurity at the first site. Therefore, the impurities are defined as
$\widetilde \epsilon_{0} = \widetilde \epsilon$ and 
$\widetilde \epsilon_{1} = \widetilde \chi |C_1|^{2}$ with 
$\widetilde \chi = \chi \sqrt{K}$ 
and $\widetilde \epsilon= \epsilon \sqrt{K}$.

Now, following the same approach as in the case of a one-dimensional chain,
we obtain all possible symmetric, antisymmetric and asymmetric SL states
for various values of the impurity strengths. 
The full phase diagram of SL states in the plane of $\epsilon$ versus $\chi$ 
for the Cayley tree ($K=4$) with a linear impurity and a nonlinear impurity 
embedded in it is shown in Fig.~7.  
We note that the maximum number of SL states in this case is 7. 
Thus, a linear impurity and a nonlinear impurity play the same role as that of
the two nonlinear impurities as long as the maximum number of SL states is 
concerned.
However, the phase diagram is completely different from that for
a Cayley tree with two nonlinear impurities. 
The one-by-one increment of the number of SL states from one region
to another is retained.

\begin{center}
\epsfig{file=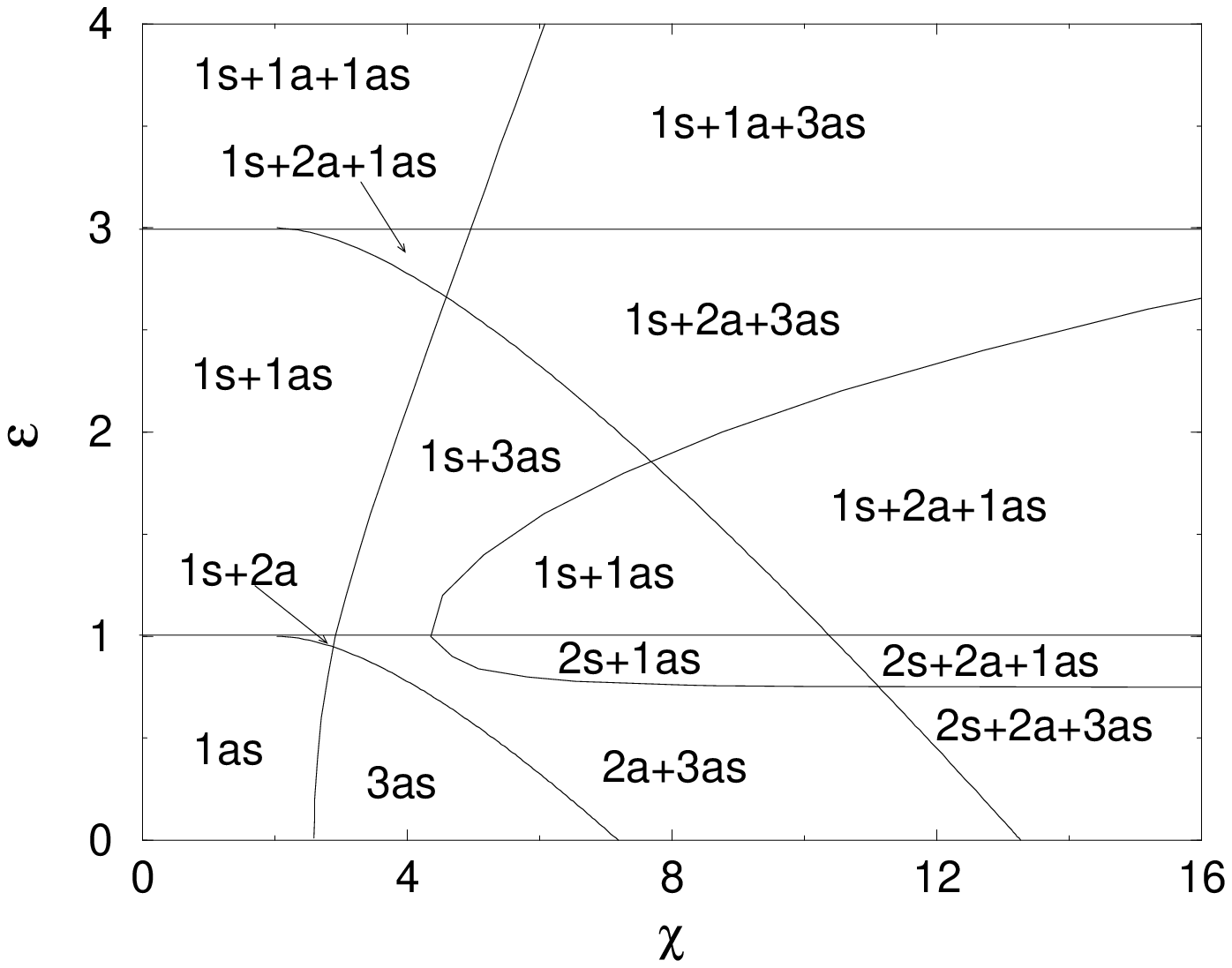,width=7.8cm,height=6.5cm}
\begin{figure}[bh]
\caption{Phase diagram for SL states in the Cayley tree
         of connectivity $K=2$ with a linear and a nonlinear impurity
         embedded in it.}
\label{fig7}
\end{figure}
\end{center}

\section{summary}
Formation of SL states due to the presence of linear and nonlinear 
impurities is studied. 
Three different cases are considered. 
First, a one-dimensionl chain with a linear impurity and a 
nonlinear impurity is considered, 
and the maximum number of SL states is found to be five.
If one considers that the maximum number of SL states due to 
two linear impurities was found to be two and that due to 
two nonlinear impurities was five, one may conclude that the presence 
of one linear impurity and one nonlinear impurity plays the same role
as that of two nonlinear impurities as long as the maximum number of
SL states is concered. 
Variation of the energy of SL states as a function of nonlinear strength
is studied and it is found that when a pair of states of the same type
appear at the same value of $\chi$, 
one of them goes away from the band edge and another approaches the 
band edge as $\chi$ increases. 
Second, a linear chain with linear and nonlinear site 
energies at alternate sites is considered and a rich phase diagrams
for SL states are obtained. 
The maximum number of SL states are found to be 9,
which is larger than that for any system studied so far in this direction. 
These SL states may be called self-localized states because 
they appear even though the translational symmetry in the system is preserved. 
Third, as for the case of a different lattice structure, 
the Cayley tree with a linear impurity and a nonlinear impurity 
is considered and a rich phase diagram of SL states is obtained. 
The maximum number of SL states is found to be 7.

\acknowledgments
This work was supported by the Korea Research Foundation Grant
(KRF-2000-015-DP0101). The authors are grateful for the support.


\begin{thebibliography}{99}
\bibitem{scot} 
  J. C. Eilbeck, P. S. Lomdahl and A. C. Scott, Physica D {\bf 16},
  318 (1985).
\bibitem{pete} 
  P. L. Christiansen and A. C. Scott, {\it Devydov's Soliton Revisited: 
  Self-Trapping od Vibrational Energy in Protein} 
  (Nato ASI Series B: Physics, vol. 243, New York: Plenum (1991)). 
\bibitem{emin} 
  D. Emin, Phys. Today {\bf 35}, 34 (1982).
\bibitem{camp} 
  D. K. Campbell, A. R. Bishop and K. Fesser, Phys. Rev. B {\bf 26},
  6862 (1982).
\bibitem{chen} 
  D. Chen, M. I. Molina and G. P. Tsironis, J. Phys. Conden. Matt. {\bf 5}, 
  8689 (1993).
\bibitem{wany} 
  Y. Wan and C. M. Soukoulis, Phys. Rev. B {\bf 40}, 12264 (1989).	
\bibitem{mol1} 
  M. I. Molina and G. P. Tsironis, Phys. Rev. Lett. {\bf 73}, 464 (1994).
\bibitem{cohe}   
  M. H. Cohen, E. N. Economou and C. M. Soukoulis, Phys. Rev. Lett. {\bf 51},
  1202 (1983).
\bibitem{mill}
  W. Chen and D. L. Mills, Phys. Rev. Lett. {\bf 58}, 160 (1987).	       
\bibitem{abri} 
  A. A. Abrikosov, A. I. Buzdin and M. L. Kulic, Supercond.  Sci. Technol. 
  {\bf 1}, 260 (1989).
\bibitem{daux} 
  T. Dauxois, M. Peyrard and A. R. Bishop, Phys. Rev. E {\bf 47}, 684 (1993).
\bibitem{caid} 
  D. Cai, A. R. Bishop and N. Gronbech-Jensen, Phys. Rev.  Lett. 
  {\bf 72}, 591 (1994).
\bibitem{mol2} 
  M. I. Molina, G. P. Tsironis and D. Hennig, Phys. Rev. E {\bf 50}, 
  2365 (1994).
\bibitem{mol3}
  M. I. Molina and G. P. Tsironis, Int. J. Mod. Phys. B {\bf 9}, 1899 (1995).
\bibitem{hui1} 
  Y. Y. Yui, K. M. Ng and P. M. Hui, Phys. Lett. A {\bf 200}, 325 (1995).
\bibitem{hui2}
  Y. Y. Yui, K. M. Ng and P. M. Hui, Solid State Commun. {\bf 95}, 801 (1995).
\bibitem{hui3} 
  Y. Y. Yui, K. M. Ng and P. M. Hui, J. Phys. Conden. Matt. {\bf 8},
  2011 (1996).
\bibitem{acev} 
  A. B. Aceves, C. De Angelis, T. Peschel, R. Muschall,
  F. Lederer, S. Trillo and S. Wabnitz, Phys. Rev. E {\bf 53}, 1172 (1996).
\bibitem{wein} 
  B. Malomed and M. I. Weinstein, Phys. Lett. A {\bf 220}, 91 (1996).
\bibitem{bik1} 
  B. C. Gupta and K. Kundu, Phys. Rev. B {\bf 55}, 894 (1997).
\bibitem{bik2} 
  B. C. Gupta and K. Kundu, Phys. Rev. B {\bf 55}, 11033 (1997).
\bibitem{kun}
  B. C. Gupta and K. Kundu, Phys. Lett. A {\bf 235}, 176 (1997).
\bibitem{bik3} 
  A. Ghosh, B. C. Gupta and K. Kundu, J. Phys. Conden. Matt. {\bf 10},
  2701 (1998).
\bibitem{bik4} 
  K. Kundu and B. C. Gupta, Eur. Phys. J. B {\bf 3}, 23 (1998).
\bibitem{econ}
  E. N. Economou, {\it Green's Function in Quantum Physics}
  (Springer, Berlin, 1979).
\end{thebibliography}
\end{document}